# Optimal Low-Latency Network Topologies for Cluster Performance Enhancement

Yuefan Deng, *Member, IEEE*, Meng Guo, Alexandre F. Ramos,
Xiaolong Huang, Zhipeng Xu and Weifeng Liu

**Abstract**—We propose that clusters interconnected with network topologies having minimal mean path length will increase their overall performance for a variety of applications. We approach our heuristic by constructing clusters of up to 36 nodes having Dragonfly, torus, ring, Chvatal, Wagner, Bidiakis and several other topologies with minimal mean path lengths and by simulating the performance of 256-node clusters with the same network topologies. The optimal (or sub-optimal) low-latency network topologies are found by minimizing the mean path length of regular graphs. The selected topologies are benchmarked using ping-pong messaging, the MPI collective communications, and the standard parallel applications including effective bandwidth, FFTE, Graph 500 and NAS parallel benchmarks. We established strong correlations between the clusters' performances and the network topologies, especially the mean path lengths, for a wide range of applications. In communication-intensive benchmarks, clusters with optimal network topologies out-perform those with mainstream topologies by several folds. It is striking that a mere adjustment of the network topology suffices to reclaim performance from the same computing hardware.

———————————— ◆ ————————————

## 1 INTRODUCTION

The ever increasing peak processing speeds of supercomputers — culminating at IBM Summit [1] with its peak speed of 201 PFlops, and 2,397,824 cores — brings exascale era within reach by systems and applications developers. For achieving the milestone of exascale computing, the developers must reduce power consumption and increase processing speeds by means of, *e.g.*, design of power-efficient processors (and other components) capable of delivering higher local

———————————————————

- *Y. Deng is with the Department of Applied Mathematics and Statistics, Stony Brook University, Stony Brook, NY 11794, USA and Shandong Computer Science Center (National Supercomputer Center in Jinan), Jinan, Shandong 250101, P.R. China.
  E-mail: Yuefan.Deng@StonyBrook.edu.*
- *M. Guo is with Shandong Computer Science Center (National Supercomputer Center in Jinan) and Shandong Provincial Key Laboratory of Computer Networks, Qilu University of Technology (Shandong Academy of Sciences), Jinan, Shandong 250101, P.R. China.
  E-mail: guomeng@sdas.org.*
- *A. F. Ramos is with Escola de Artes, Ciências e Humanidades, Núcleo de Estudos Interdisciplinares em Sistemas Complexos, Center for Translational Research in Oncology (LIM-24), Instituto do Câncer do Estado de São Paulo, Faculdade de Medicina, Universidade de São Paulo, Avenida Arlindo Béttio 1000, CEP 03828-000, São Paulo, São Paulo, Brazil and Shandong Computer Science Center (National Supercomputer Center in Jinan), Jinan, Shandong 250101, P.R. China.
  E-mail: Alex.Ramos@usp.br.*
- *X. Huang is with the Department of Applied Mathematics and Statistics, Stony Brook University, Stony Brook, NY 11794, USA and Shandong Computer Science Center (National Supercomputer Center in Jinan), Jinan, Shandong 250101, P.R. China.
  E-mail: Xiaolong.Huang@StonyBrook.edu.*
- *Z. Xu is with the School of Data and Computer Science, Sun Yat-sen University, Guangzhou, Guangdong 510006, P.R. China and the Department of Applied Mathematics and Statistics, Stony Brook University, Stony Brook, NY 11794, USA.
  E-mail: xuzhp9@mail2.sysu.edu.cn.*
- *W. Liu is with the School of Information Science and Engineering, University of Jinan, Shandong, 250022, P.R. China.
  E-mail: ise_liuwf@ujn.edu.cn.*

performance and design of networks for system performance capable of delivering low-latency and high-bandwidth communications. Those goals have been incrementally achieved, as indicated by the IBM Summit having higher ratio of performance to power consumption than that of the Taihu Light; the smaller number of cores delivering such faster processing; or on the changes of the top positions of the Top500 lists of June and November of 2018 [1] when Sierra machine surpassed Taihu Light with a new HPL result. Performance increase, however, cannot rely only on raising individual processors clock speed because of the power wall of the Moore's law [2]. Hence, interconnecting millions of processors in optimally designed networks, especially the network topologies having minimal latency, is the key for further increase of supercomputers' peak speeds. An additional requirement for network optimality is to provide a consistent environment to enable new programming models to extract more performance. To attend that, theoretical insights [3, 4] for describing, designing, analyzing, and optimizing the next-generation interconnection networks for increasing global processing speeds of supercomputers may become a major tool for the HPC community.

In this manuscript, we approach the problem of enhancing a cluster's performance using symmetric and engineeringly feasible minimal latency network topologies supported by a new framework for designing regular graphs of degree $k$ with rotational symmetry and minimal mean path length. The graphs support the network topologies of the directly connected clusters that we benchmarked. The optimal graphs enabled building a cluster which may outperform a torus of the same degree by a factor of up to 3. Our graphs of degree 3 can achieve the same performance of the torus of degree 4 — a clear reduction of hardware costs, engineering complexity, and power consumption. Our results showing the favorable impact of optimal graphs on performance open a new avenue of theoretical and experimental research for supercomputer architects. Related work is discussed in Section 2 and Section 3 presents our algorithm for designing a network topology and the cluster that we used on our analysis. Section 4 presents and examines graph properties supporting different clusters designs and their benchmark results. Concluding remarks are presented in Section 5.

## 2 RELATED WORK

Although we do not intend to do a comprehensive review on the extensive literature concerning supercomputers and data centers, a brief discussion on the potential use of our approach in complement to existing technologies is given. Despite active theoretical investigations on designing network topologies for clusters [5, 6] their use in actual machines has not been a priority since the early days of parallel computing [7] because of potential engineering complications and lack of a measure of performance gains. Network topologies are the main elements affecting supercomputer interconnection network performance, and for decades, tori of 3D through 6D [8-14], hypercubes of various dimensions [15-17], fat-trees [18-20] and off-the-shelf Ethernet or adapted InfiniBand switched fabrics have been the mainstream network subsystems.

In general, the system architecture aims at providing maximal connectivity, scalable performance, minimal engineering complexity, and least monetary cost [3]. An ideal network of a fixed node degree must satisfy performance requirements including small network diameter, broad bisection width, simple symmetric topology, engineering feasibility, and modular, expandable design [3]. For example, mesh topology has low node degree and engineering complexity, but its large network diameter and average distance dampen node-to-node communications; fat-tree realizes the maximum bisection width with large diameter; the torus and its derivative k-ary n-cube [21] has lower node degree, diameter, and average distance. Hybrid 6D mesh/torus TOFU interconnect is incorporated in K computer [14], modified 3D torus with combined 2-node is designed to form the Cray Gemini interconnect [13], upgrading from the traditional 3D torus topology as in Cray SeaStar [8, 9], IBM Blue Gene/L [10], Blue Gene/P [11] and 5D torus as in IBM Blue Gene/Q [12]. Other variants of torus such as the SRT [22] and RDT [23] networks, variant of k-ary n-cube such as the Express Cubes [24] and interlaced bypass torus (iBT) [25, 26] use the technique of adding bypass links. Modifications of fat-tree [27, 28] have also been done to reduce its complexity and cost. Recently, high-radix topologies such as Dragonfly [29] and Aries interconnect [30] have been thoroughly studied and implemented as well. However, to the best of our knowledge network topologies aiming at minimizing mean path length have not been used in supercomputers architecture yet.

On the other hand, use of data centers for cloud computing has been rapidly increasing, and challenges architects to build machines which amounts of processing nodes, memory, and switches grow steadily while keeping the machine operational. That poses scalability and fault detection, along with maximal bisection bandwidth, as key features of Data Center Networks (DCN). Instead of reaching that by addition of switch layers, recent advances propose the use of optical networks for switchers interconnects enabling switch to switch communication while replacing bigger top-of-rack switchers [31]. That approach may be complemented by ours by constructing optimal networks of switches with reduced latency. In that case, there will be two optimization procedures, for minimizing MPL and labeling pairs of communicating optical channels, which will enable the small network of switches to perform optimally under constraint of a finite numbers of ports. Symmetry of our optimal network topologies enable low levels of engineering complexity, as exemplified by our prototype machines.

## 3 DISCOVERY OF OPTIMAL NETWORK TOPOLOGIES AND CLUSTER DESCRIPTION

We assume that a cluster of $N$ nodes can have its processing speeds increased if we design a network topology of minimal latency. Hence, we propose a new algorithm to discover minimal MPL symmetric graphs to support optimal low-latency network topologies for clusters and test our proposition on a directly connected cluster.

### 3.1. Discovery of Optimal Network Topologies

To obtain optimal network topologies, we search for $N$-vertex degree-$k$ regular graphs, denoted by $(N, k)$, with minimal mean path length (MPL). Cerf et al. first calculated the lower bound of MPL for any regular graph [32], and discovered small degree-3 graphs with up to 24 vertices whose MPL is minimal. Additionally, it was proved that the diameters of such optimal graphs are also minimal. The exhaustive computer search of an optimal graph of fixed size and degree is computationally expensive, e.g., the number of non-isomorphic 32-vertex degree-3 regular graphs, labeled as (32,3), is ~$10^{13}$ [34]. Thus, heuristics have been developed using greedy local search [35], simulated annealing [36], or theoretical graph product and construction [37] for reduced search duration.

For finding the graphs, we implemented the graph parallel exhaustive search using the enumeration algorithms *snarkhunter* [38, 39] and *genreg* [40], with builtin split option for parallelization and girth (the length of the smallest cycle in the graph) option as constraint. Optimal graphs having large girths [33] help reduce the search space, e.g., a reduction from ~$10^{13}$ non-isomorphic (32,3) regular graphs (with no girth constraint) to ~$10^5$ by a constraint of girth $\geq 7$ [34]. This method was used for finding the (32,3)-optimal graph. However, the cost of exhaustive search with more vertices or higher degree is astronomical even under girth constraint.

To find larger optimal graphs, we relied on heuristics based on simulated annealing [41] and edge swap. Starting from a random Hamiltonian regular graph, i.e., with an

 

embedded ring, we perform the edge swap on non-ring edges.

---

**Algorithm.** SA Search with Edge Swap

---

1:   Initialization:
      $G \leftarrow$ random Hamiltonian regular graph
      $T \leftarrow$ high temperature $T_{start}$
2:   **While** *not stopping criteria* **do**
3:       Generate new graph $G_{new}$ by swapping
      non-ring edges
4:       $\Delta\text{MPL} = \text{MPL}(G_{new}) - \text{MPL}(G)$
5:       **if** $\Delta\text{MPL} < 0$
6:          $G \leftarrow G_{new}$
7:       **else**
8:          Accept $G_{new}$ with probability $exp(-\Delta\text{MPL}/T)$
9:       $T \leftarrow \gamma \cdot T$

---

The stopping criteria may include falling below the prescribed lowest temperature $T_{end}$, exceeding the number of total iterations $n_{iter}$ and/or approaching the MPL theoretical lower bound. The cooling rate $\gamma$ depends on different cooling schedules. We use the classic exponential cooling scheme first proposed by Kirkpatrick et al. [41] and set $\gamma = exp(log(T_{end}/T_{start})/n_{iter})$ so that the temperature is cooled down to $T_{end}$ after $n_{iter}$ iterations. By this method, we have discovered the (32,4)-Optimal graph. It is worth mentioning that the final layout of the optimal graphs of (32,3) and (32,4) are 90° rotationally symmetric after the MPL optimization search. For each optimal graph, we re-order the vertices on the ring according to its different Hamiltonian cycles and look for rotational symmetry of these isomorphic layouts. The coloring of the edges helps to visualize this symmetric design. Fixing such symmetric structure is also one way to reduce the search space, which we apply to the optimization of larger-scale topologies.

### 3.2. Cluster Description

To verify our approach, we have constructed a switchless Beowulf cluster named "Taishan" (Fig. 1) with up to 32 nodes. Each node has eight communication ports, with two of them used for cluster management and storage. With it, we can evaluate performances of clusters with network topologies supported by graphs of degrees 2 - 6. That enabled us to benchmark the impact of network topology on performance. Because of hardware homogeneity, we conclude that our results on the impact of network topology remain valid when cutting edge technology is used.

Because of budget limits we use a low-end hardware to build a functional prototype suited for investigating the impact of the network topology on the cluster's performance. Each node of Taishan has 1 Intel Celeron 1037U dual-core processor (1.80 GHz, 2M Cache), 1×8 GB DDR3 SODIMM (1600 MHz, 1.35V), 128 GB SSD and eight Intel 82583V Gigabit Ethernet controllers (PCIe

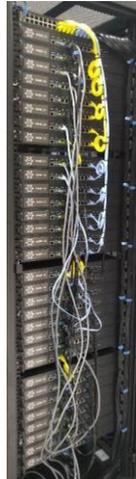

Fig. 1. Taishan Beowulf cluster.

v.1.1, 2.5 GT/s). We use CentOS Linux 6.7 (kernel 2.6.32) as operational system, and NFS for sharing files through one of the ports that is connected to a 48-port Gigabit Ethernet switch. Processes communicate directly through node's ports interconnected accordingly with the supporting graph adjacency rules. We use GCC version 4.4.7 and MPICH 3.2 for compiling and running our parallel programs. Static routing is used accordingly with Floyd's algorithm [42] to ensure the shortest path length and lowest congestion.

## 4 ANALYSIS OF GRAPH PROPERTIES AND CLUSTER BENCHMARKS

### 4.1. Comparative Analysis of Optimal Network Topologies

In order to evaluate the effects of the optimal network topologies on the cluster performance, we have designed several network topologies using regular graphs $(N, k)$ with $N = 16, 32$ and $k = 2, 3, 4$. The topologies of the benchmarked clusters of 16 nodes are: ring (R), Wagner (W) [43], Bidiakis (B) [44], 4×4 torus (T) (4D hypercube), and two optimal graphs (O) re-discovered by our parallel exhaustive search. The 32 vertices clusters used the ring, Wagner, Bidiakis, 4×8 torus, Chvatal (C) [43], and the two optimal graphs obtained by our parallel exhaustive search and simulated annealing. We also compute the bisection width (BW) of each topology using the KaHIP program, which efficiently achieve a balanced partition of a graph [45]. We refer to each cluster as $(N, k)$ −X, where X is the 1st letter of, or the name of the supporting graph. The evaluated network topologies and respective graph properties are presented in TABLE 1, while Fig. 2 shows the graphs (left) and their corresponding hop distances vs latency plots (right) obtained by actual ping-pong messaging tests. In all graphs of Fig. 2, we fit the ping-pong latency denoted by $T$ as a linear function of the hop distance, namely, $T = T_0 + \alpha \cdot h$ where $T_0$ is the network initiating time, $\alpha$ is the slope, and $h$ is the hop distance. $\rho$ is the Pearson correlation between the latency and the hop distance distributions.

TABLE 1
Graph Properties of Benchmarked Low-Radix Topologies

| Topology | D | MPL | BW | Topology | D | MPL | BW |
|---|---|---|---|---|---|---|---|
| **(16,4)-Optimal** | 3 | 1.75 | 12 | **(32,4)-Optimal** | 3 | 2.35 | 16 |
| | | | | (32,4)-Chvatal | 4 | 2.55 | 8 |
| (16,4)-Torus | 4 | 2.13 | 8 | (32,4)-Torus | 6 | 3.10 | 8 |
| **(16,3)-Optimal** | 3 | 2.20 | 6 | **(32,3)-Optimal** | 4 | 2.94 | 10 |
| (16,3)-Bidiakis | 5 | 2.53 | 4 | (32,3)-Bidiakis | 9 | 4.06 | 4 |
| (16,3)-Wagner | 4 | 2.60 | 4 | (32,3)-Wagner | 8 | 4.61 | 4 |
| (16,2)-Ring | 8 | 4.27 | 2 | (32,2)-Ring | 16 | 8.26 | 2 |

TABLE 1 shows the diameters (D), mean path length (MPL) and bisection width (BW) of the graphs supporting the benchmarked networks. For all $(N, k)$ graphs, the optimal topology has minimal MPL and D, and maximal BW. Hence, we expect that the optimal graphs will support a network topology of low latency, because of shorter MPL and D (see ping-pong test results in Fig. 2), and high throughput, because of larger BW [46]. Indeed, results present in next section lead to similar conclusions despite influence of communication patterns, internal algorithms, message sizes, memory access, and routing.

## 4.2. Benchmark Results and Analysis

The following representative benchmark programs were used to evaluate performance of the clusters: custom ping-pong and MPI collective communications; effective bandwidth (b_eff) [47, 48]; FFTE [49, 50]; Graph 500 [51, 52]; and the NAS Parallel Benchmarks (NPB) [53, 54]. Ping-pong tests report absolute runtime and produce a node-to-node latency matrix for each topology (see Fig 2) that is used to show correlation with supporting graph's hop distances. The remaining evaluation is done by means of the ratio of the performance speed (reciprocal of the absolute runtime) of each network topology to the corresponding ring. That generates a scatter plot of the performance ratio at y-axis versus the topology's MPL at x-axis for each benchmark, as shown in Figs. 3-9. Error bars are calculated by repeated experiments (except ping-pong and effective bandwidth). Red (or blue) points indicate the data for degree-3 (or -4) clusters. Different data points symbols represent different sub-tests of one application.

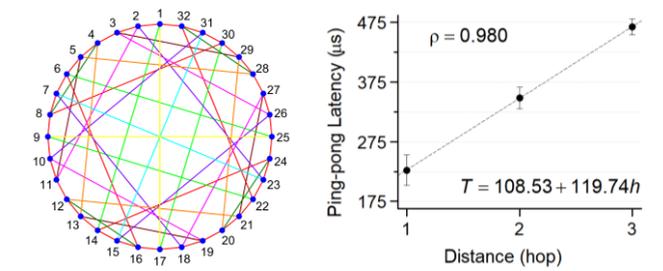

(a) (32,4)-Optimal

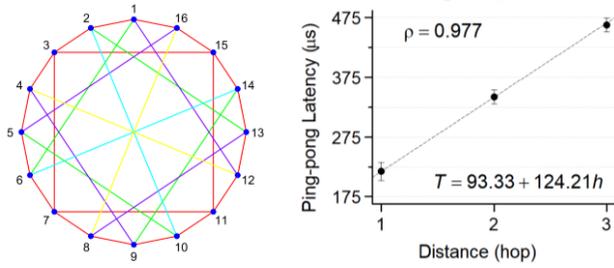

(b) (16,4)-Optimal

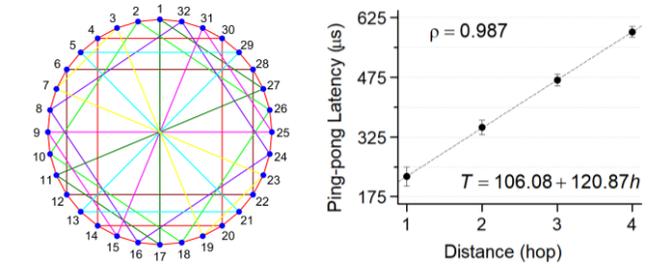

(c) (32,4)-Chvatal

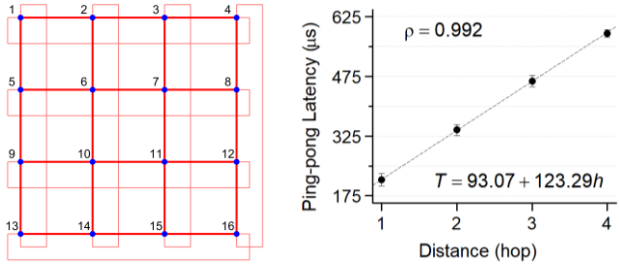

(d) (16,4)-Torus

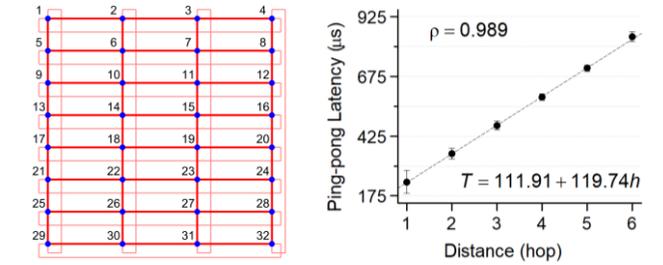

(e) (32,4)-Torus

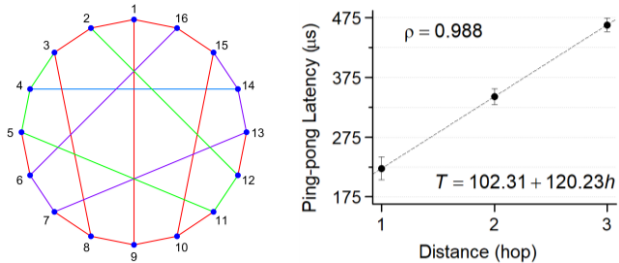

(f) (16,3)-Optimal

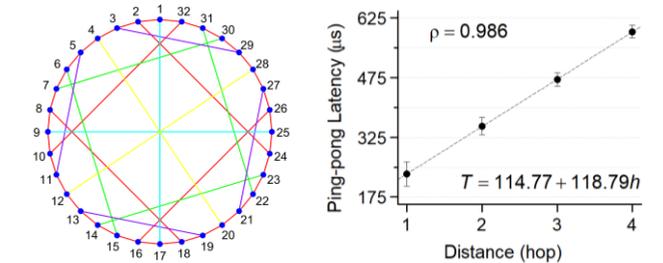

(g) (32,3)-Optimal

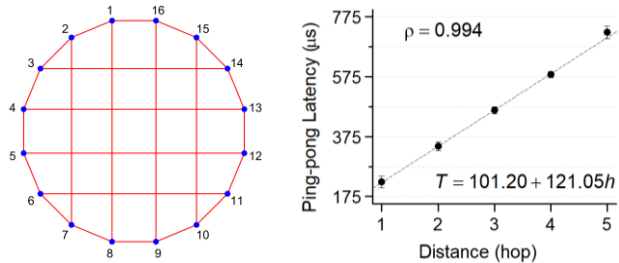

(h) (16,3)-Bidiakis

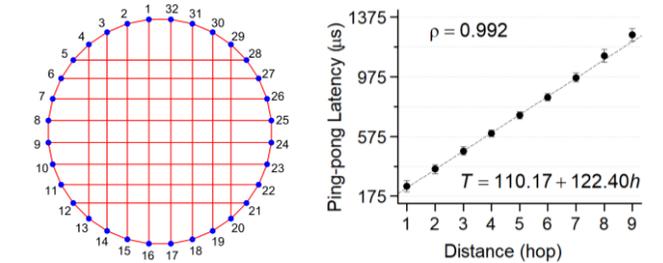

(i) (32,3)-Bidiakis

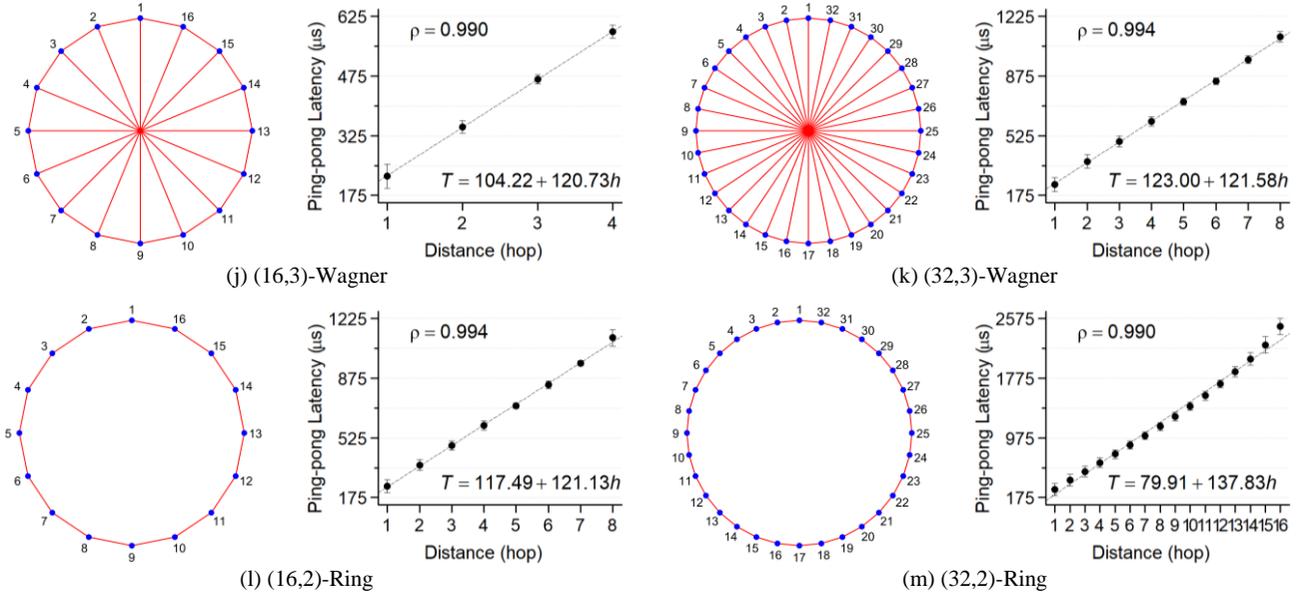

Fig. 2. Benchmarked low-radix topologies (left) and their node-to-node ping-pong latency vs. hop distance (right).

### 4.2.1. Ping-pong Tests

The routing algorithm and communication properties of the cluster in comparison to the supporting graph path lengths are evaluated by means of the ping-pong test designed using MPI_Send and MPI_Recv, with message sizes ranging from 1 byte to $2^{13}$ bytes (8 KB). Latency is measured as the average round-trip time for a message to travel between source and destination over multiple runs. We select 1 KB as the message size to output the corresponding node-to-node latency in the form of a matrix. The Pearson correlation and linear regression between node-to-node latency and hop distance were calculated for each topology as in Fig. 2, while performance ratios of average latency between all pairs of nodes for each topology are plotted in **Erro! Fonte de referência não encontrada.**.

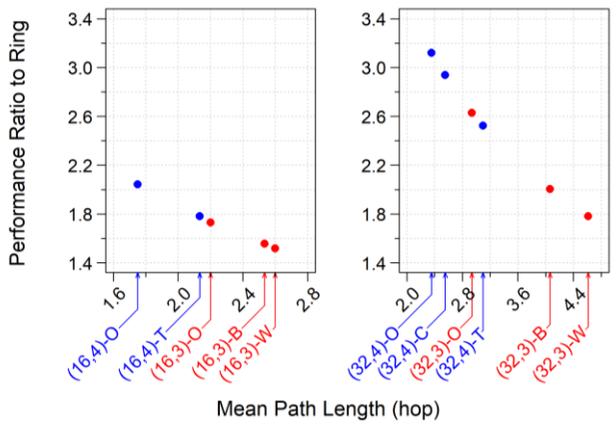

Fig. 3. Performance ratios on ping-pong tests.

Fig. 2 shows that the Pearson correlation coefficients ($\rho$) between ping-pong latency and hop distance under the shortest path routing are all greater than 0.977. Such a strong correlation is reflected on the approximately linear dependence between node-to-node latency in the network and graph's distance (hop) as indicated by the dashed line. Notice that besides (32,2)-Ring the fitting equations describing the linear relation are very similar, independently of the cluster's sizes and topologies, the average of which being $T = 107.17 + 121.15h$. (Because of the high diameter of (32,2)-Ring, message traverse and serialization start to affect the latency for long-distance transfer). Moreover, performance of ping-pong for different topologies is strongly inversely proportional to their MPL as shown in **Erro! Fonte de referência não encontrada.**. Those results also hold for larger messages of sizes up to 8 KB.

### 4.2.2. Collective Communications

Collective communications benchmarks test the performance of MPI_Bcast, MPI_Reduce (with reduce operation MPI_SUM), MPI_Scatter, and MPI_Alltoall. We choose unit messages of 1 MB and 32 MB under the constraint of 8 GB RAM available per node. On each node, the transfer message sizes are either equal to the unit message sizes or unit sizes multiplied by the number of nodes, depending on whether it is the root node, and on the MPI collective function.

MPI_Bcast, MPI_Reduce and MPI_Scatter were run multiple times with all nodes being root multiple times. Then we average the runtime over all root nodes, and then over all tests. The runtime of each test is the maximum elapsed wall clock time on all nodes. For MPI_Alltoall we conduct the test multiple times and average the runtime over all tests. The runtime of each test is the average elapsed wall clock time on all nodes.

The performance ratios to ring are plotted in Fig. 4. Collective communications are influenced by MPL, BW, traffic pattern, MPI internal algorithm, message size and memory access. For example, Wagner topology has greater MPL but shorter diameter than Bidiakis, while they have the same bisection width (TABLE 1). The shorter diameter of Wagner graph is especially pronounced in the 1 MB message MPI_Bcast (Fig. 4a) which leads to a 17% and 11% performance gain, respectively, for (16,3) and (32,3)-Wagner over Bidiakis. However, for larger messages and other MPI collective functions with similar traffic pattern such as MPI_Scatter (Fig. 4c), MPL becomes a more dominant factor and Bidiakis outperforms or at least performs equally as Wagner with slight fluctuation. Static shortest-path routing also affects the performance of collective communications. For example, torus has relatively low performance in MPI collective functions with large message, except MPI_Reduce (Fig. 4b). The low performance when transferring large message may be caused by network congestion due to static routing especially for torus, while the internal algorithm of MPI_Reduce overcomes such congestion.

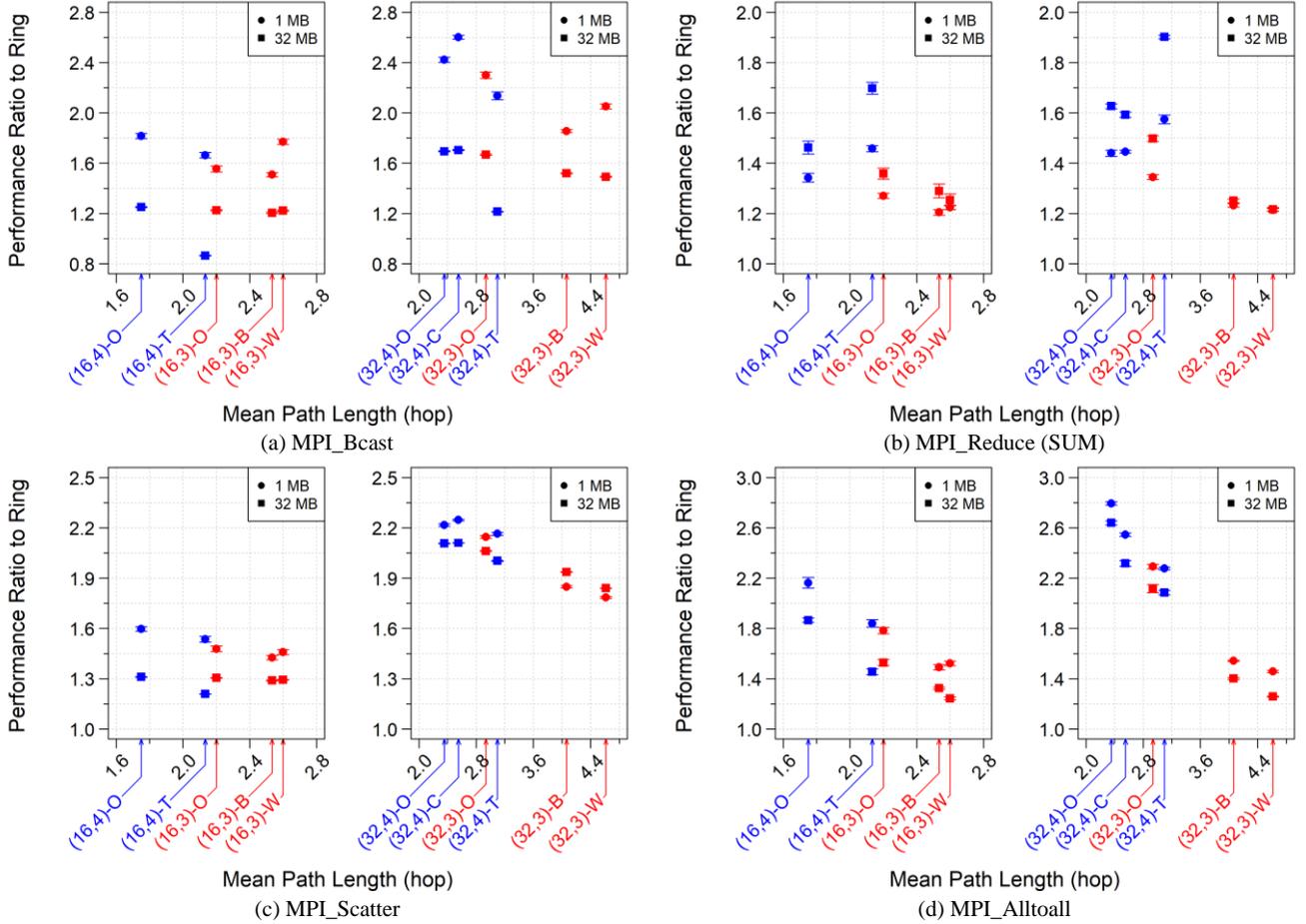

Fig. 4. Performance ratios on collective communications.

Nonetheless, despite fluctuations caused by other influential factors, the performance of collective communications has weak or strong inversely proportional relation to MPL. Particularly, MPI_Alltoall stresses global communication in the network, and shows a performance that has the strongest inversely proportional relation to MPL (Fig. 4d). The superiority of optimal topologies is demonstrated in MPI_Alltoall, where (16,4) and (32,4)-Optimal have top performance ratios of 2.16/1.87 and 2.79/2.64 to ring respectively for unit message sizes 1 MB/32 MB, an increase of 42%/50% and 92%/110% over (16,3) and (32,3)-Wagner that have the lowest ratios respectively (excluding (16,3)-Bidiakis with slight fluctuation). The role of MPL for optimizing MPI_Alltoall communication is reinforced when we compare (32,3)-Optimal and (32,4)-Chvatal, whose bisection width are, respectively, 10 and 8, with Chvatal delivering a better performance because of its smaller MPL.

### 4.2.3. Effective Bandwidth

Effective bandwidth (b_eff, version 3.6.0.1) [47] measures the accumulated network bandwidth by means of multiple communication patterns (ordered naturally and randomly) with messages of 21 sizes ranging from 1 byte to 1/128 of memory per processor, 64 MB in Taishan. It uses MPI_Sendrecv, MPI_Alltoallv and non-blocking MPI_Irecv and MPI_Isend with MPI_Waitall. The output is the average bandwidth over ring and random patterns and 21 message sizes after taking the maximum bandwidth of the three MPI methods in each measurement [48].

The performance ratios to ring are plotted in Fig. 5. A strong impact of MPL on b_eff benchmark is shown, though traffic patterns, message sizes and MPI methods may also affect performance. Indeed (16,4) and (32,4)-Optimal have the highest effective bandwidths, 686.51 MB/s (and 1066.80 MB/s), a performance gain of 38% (and 68%) over (16,3) and (32,3)-Wagner. Indeed, we can consider that performance of b_eff has an inversely proportional relation to MPL if we neglect the torus because the static shortest-path routing causes congestion in collective MPI functions.

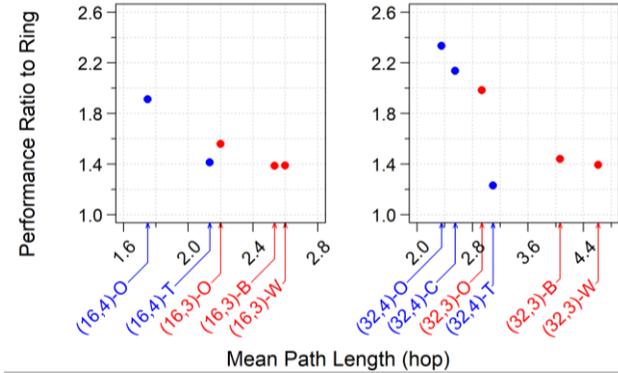

Fig. 5. Performance ratios on effective bandwidth.

### 4.2.4. FFTE

We benchmarked the version 6.0 of the parallel FFTE [49, 50] from the HPC Challenge [55, 56], which in cache-based processors [57], has data transpositions as its main bottleneck because of all-to-all communications. We perform the parallel 1D FFTE routine with transform array lengths ranging from $2^{10}$ to $2^{27}$, limited by local 8 GB RAM. Then we select $2^{21}$ and $2^{27}$ as the transform array lengths (equal to 32 MB and 2 GB in total transform array sizes).

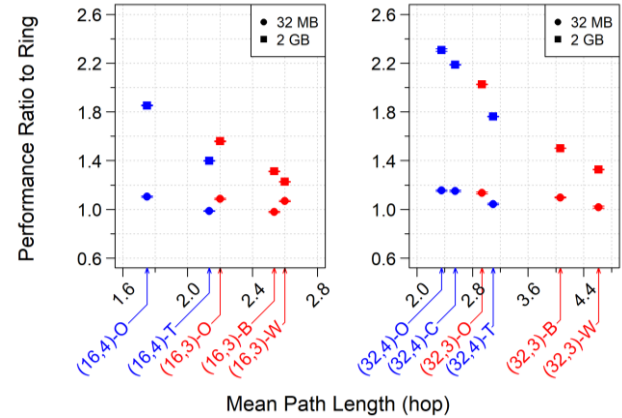

Fig. 6. Performance ratios on 1D FFTE.

**Erro! Fonte de referência não encontrada.** shows the performance plots of 1D FFTE. Transforming larger arrays stresses the network such that 1D FFTE performs with almost linear dependence of MPL. When transforming 2 GB array in 1D FFTE, (16,4) and (32,4)-Optimal topologies have top performance ratios of 1.85 and 2.31 to ring, a gain of 51% and 74% over (16,3) and (32,3)-Wagner. For arrays < 32 MB the performances are almost uniform for all network topologies.

### 4.2.5. Graph 500

The Graph 500 (version 3.0.0) [51, 52] tests large-scale graph algorithms, where multiple breadth-first search (BFS) and single-source shortest path (SSSP) computations are performed on an extremely large undirected graph generated and distributed in the beginning of the test. Graph 500 evaluates data-intensive performance in supercomputers reporting the mean TEPS (traversed edges per second). The best choice for test scale accordingly with local RAM was 27, generating an initial unweighted graph of 24 GB for BFS, and an initial weighted graph of 40 GB for SSSP.

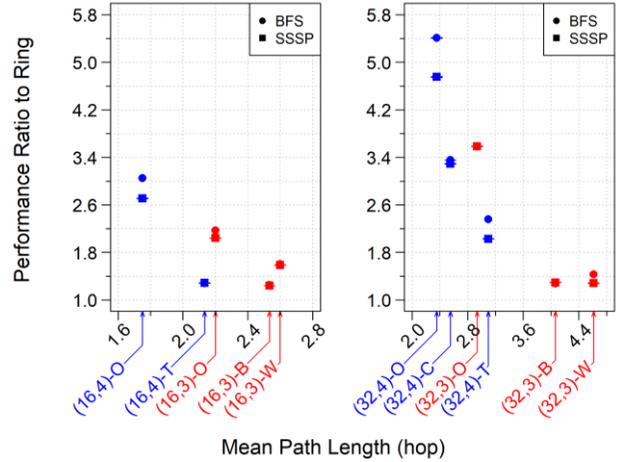

Fig. 7. Performance ratios on Graph 500.

Fig. 7 shows the performance of Graph 500 benchmark. A strong inversely proportional relation to MPL is exhibited, despite fluctuations on torus (because of congestion), Bidiakis and (32,4)-Chvatal. The relatively high diameter of Bidiakis compared with Wagner and relatively low bisection width of (32,4)-Chvatal compared with (32,3)-Optimal topology (TABLE 1) weaken their performances as well. However, MPL keeps playing a major role on Graph 500 with (16,4) and (32,4)-Optimal having top performances of, respectively, 3.05/2.71 and 5.41/4.75 for BFS/SSSP, a gain of 90%/71% and 278%/271% over (16,3) and (32,3)-Wagner.

### 4.2.6. NAS Parallel Benchmarks (NPB)

The NAS Parallel Benchmarks (NPB version 3.3.1 on MPI) [53, 54] contain a set of programs derived from computational fluid dynamics (CFD) applications, with built-in run time reporting. We run integer sort (IS), conjugate gradient method (CG) for approximating the smallest eigenvalue, multi-grid solver (MG) for 3D Poisson PDE, FFT solver (FT) for 3D PDE NPB kernels, and lower-upper (LU) Gauss-Seidel solver pseudo-application [58]. IS uses intensive data communication, while also testing random memory access and integer computation speed; CG tests unstructured long-distance communication and irregular memory access; MG tests highly structured short- and long-distance communication with intensive memory access; FT tests long-distance all-to-all communication [53, 54, 58]. For each benchmark, we choose the standard problem sizes: Class A, B, and C because of local memory constraints.

The performance ratios to ring for Classes A and C are shown in Fig. 8. Note that traffic patterns, internal algorithms, problem sizes, memory access and static shortest-path routing, apart from MPL and BW, affect the performance of NPB. The performances of CG (Fig. 8b) and MG (Fig. 8c) are similar to MPI_Reduce (Fig. 4b), in which torus shows relatively high performance. In these benchmarks, the static routing for torus does not cause congestion with internal algorithms and memory access benefitting the torus. LU (Fig. 8e) shows a nearly uniform performance over all benchmarked topologies, a result attributable to its limited parallelism [54], i.e., low communication to computation ratio. However, NPB performance exhibits weak, or even strong, dependence on MPL as in IS (Fig. 8a) and FT (Fig. 8d) resembling, respectively, Graph 500 (Fig. 7) and 1D FFTE with 2 GB array size (**Erro! Fonte de referência não encontrada.**), as expected for benchmarks requiring heavy global communication. IS, and FT Class A/C problem sizes are $2^{23}$ / $2^{27}$ resulting in, respectively, 32 MB/512 MB total integer array sizes, and 128 MB/2 GB transform array sizes. In IS Cass A/C, (16,4) and (32,4)-Optimal topologies have top performance ratios of 2.70/2.89 and 3.89/4.32, respectively, a gain of 79%/93% and 153%/202% over (16,3) and (32,3)-Wagner. In FT Class A/C, the optimal graphs, 1.72/1.66 and 2.31/2.35, outperforms both Wagner graphs with a gain of 26%/40% and 56%/81%, respectively.

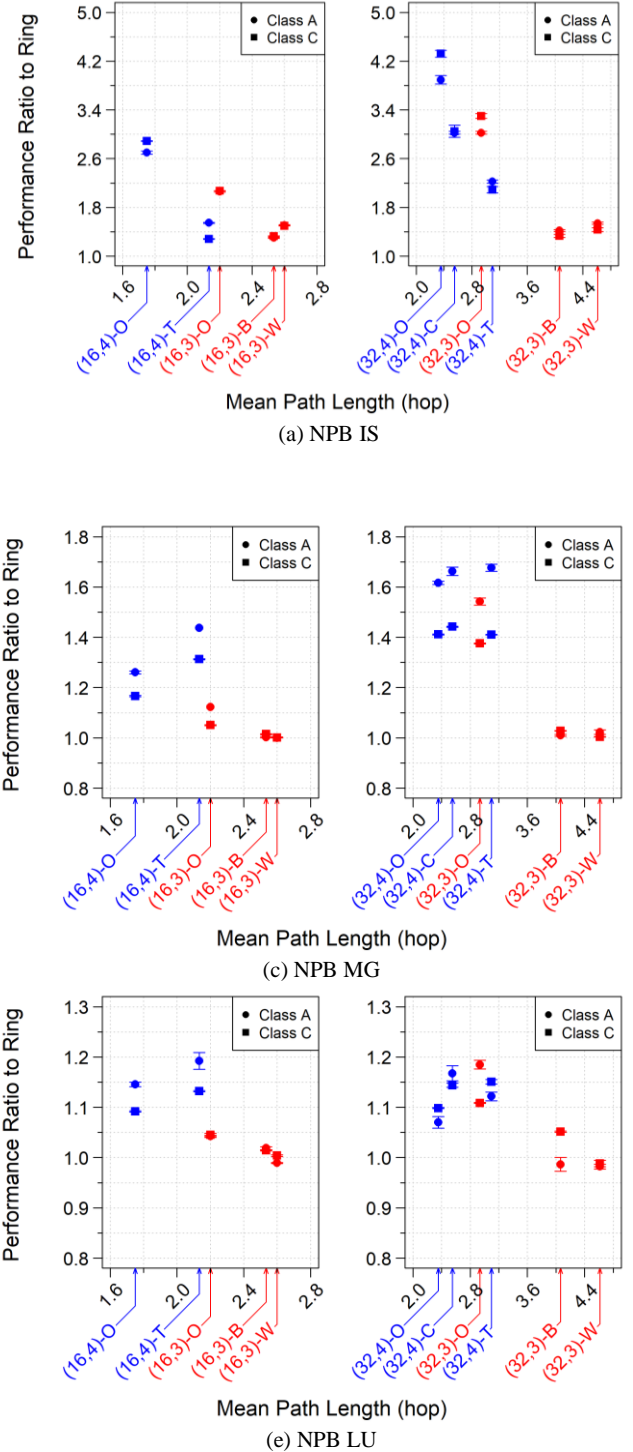

Fig. 8. Performance ratios on NPB.

## 4.3. Comparison of High-Radix Optimal Topologies and the Corresponding Dragonfly Topologies

We apply our optimization method to obtain high-radix optimal topologies to their corresponding Dragonfly topologies, see Fig. 9 and TABLE 2. The high-radix optimal topologies are obtained using methods of Section 3 added by random iteration of Hamiltonian graphs with rotational symmetry. The optimal topologies have minimum diameter (D), MPL and higher bisection width (BW) than Dragonfly. The benchmarks are performed using Taishan Beowulf cluster, but we exclude FFTE and NPB as they require the total number of nodes to be power of two. TABLE 3 shows the performance ratios of optimal topologies over Dragonfly topologies on effective bandwidth, Graph 500 and MPI_Alltoall.

The results reveal that optimal topologies with the lowest MPL have better performances over corresponding Dragonfly topologies, especially on applications such as Graph 500 and MPI_Alltoall with large amount of global communication. In MPI_Alltoall with unit message sizes 1 MB/32 MB, (30,5)-Optimal has top performance increase of 67%/80% over Dragonfly. For other MPI collective functions, the performances of optimal and corresponding Dragonfly topologies are very similar, due to their equal diameter and relatively close MPL.

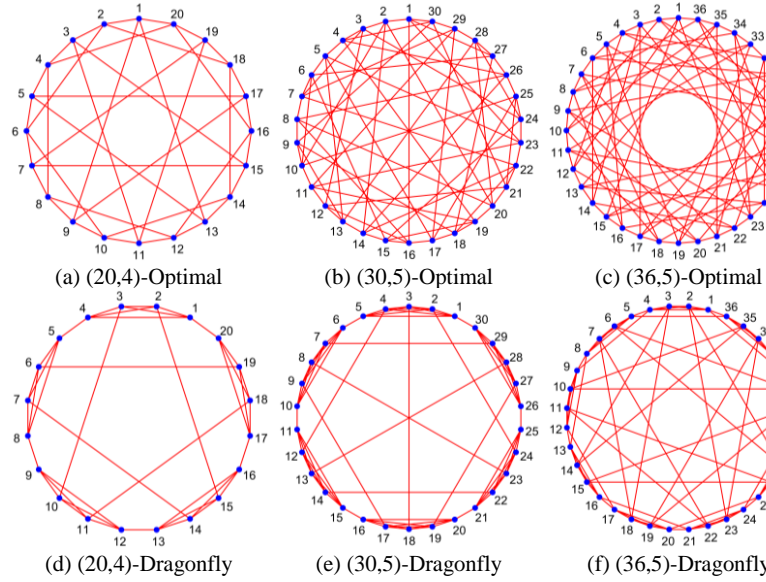

(a) (20,4)-Optimal  (b) (30,5)-Optimal  (c) (36,5)-Optimal

(d) (20,4)-Dragonfly  (e) (30,5)-Dragonfly  (f) (36,5)-Dragonfly

Fig. 9. The high-radix optimal topologies (top) and the corresponding Dragonfly topologies (bottom).

### TABLE 2 Graph Properties of the Optimal Topologies and the Corresponding Dragonfly Topologies

| Topology | D | MPL | BW |
|---|---|---|---|
| **(20,4)-Optimal** | 3 | **1.95** | **10** |
| (20,4)-Dragonfly | 3 | 2.26 | 8 |
| **(30,5)-Optimal** | 3 | **1.97** | **25** |
| (30,5)-Dragonfly | 3 | 2.38 | 9 |
| **(36,5)-Optimal** | 3 | **2.14** | **24** |
| (36,5)-Dragonfly | 3 | 2.34 | 20 |

### TABLE 3 Relative Performances of the Optimal Topologies Over the Corresponding Dragonfly Topologies

| Topology $(N, k)$ | b_eff | Graph 500 | | MPI_Alltoall | |
|---|---|---|---|---|---|
| | | BFS | SSSP | 1 MB | 32 MB |
| (20,4) | 1.09 | 1.14 | 1.16 | 1.22 | 1.46 |
| (30,5) | 1.14 | 1.23 | 1.20 | 1.67 | 1.80 |
| (36,5) | 1.02 | 1.03 | 1.02 | 1.15 | 1.32 |

## 4.4. Large-Scale Topology Optimization and Simulation Analysis

### 4.4.1. Comparative Analysis of Larger-Scale (Sub)Optimal Network Topologies

We obtain the suboptimal topologies of 256 nodes and degrees 3, 4, 6, 8 and optimal topologies of 252 and 264 nodes and degree 11 using random iteration of Hamiltonian graphs with rotational symmetry. The low-radix suboptimal topologies are compared with topologies of the same sizes and degrees: ring, Wagner, Bidiakis, 16×16 torus (4D hypercube), 4×8×8 torus and 4×4×4×4 torus (8D hypercube), as shown in TABLE 4. The high-radix optimal topologies are compared with corresponding Dragonfly topologies, as shown in **Erro! Fonte de referência não encontrada.**. Some of the topology figures are listed in Appendix A. For the optimal topologies, we further select the one with minimal 1D total cable length of Hamiltonian layout, due to the large number of random optimal topologies discovered. For the suboptimal topologies, we also calculate their gaps of diameter and MPL compared to the theoretical lower bounds respectively.

Tables TABLE 4 and **Erro! Fonte de referência não encontrada.** show that the (sub)optimal topologies have the smallest diameter (D), MPL and highest bisection width (BW) among the topologies of the same sizes and degrees. For the gaps of D and MPL of suboptimal topologies, the diameter gap is within 1 and MPL gap is within 2% compared to the theoretical lower bounds. This shows our optimization method is effective on the large scale. The current optimization runtime is 96 hours and one may further extend the runtime or improve the method to obtain better (sub)optimal topologies.

### TABLE 4 Graph Properties of Simulated Low-Radix Topologies

| Topology | D† | MPL† | BW |
|---|---|---|---|
| **(256,8)-Suboptimal** | **3+1** | **2.72+0.03** | **298** |
| (256,8)-Torus | 8 | 4.02 | 128 |
| **(256,6)-Suboptimal** | **4+0** | **3.11+0.06** | **192** |
| (256,6)-Torus | 10 | 5.02 | 64 |
| **(256,4)-Suboptimal** | **5+1** | **4.09+0.05** | **92** |
| (256,4)-Torus | 16 | 8.03 | 32 |
| **(256,3)-Suboptimal** | **7+1** | **5.59+0.08** | **46** |
| (256,3)-Bidiakis | 65 | 25.09 | 4 |
| (256,3)-Wanger | 64 | 32.62 | 4 |
| (256,2)-Ring | 128 | 64.25 | 2 |

†*The D and MPL of suboptimal topologies are written as the sum of the theoretical lower bounds and the difference to final values.*

### TABLE 5 Graph Properties of the Larger-Scale Optimal Topologies and the Corresponding Dragonfly Topologies

| Topology | D | MPL | BW |
|---|---|---|---|
| **(252,11)-Optimal** | 3 | 2.47 | 388 |
| (252,11)-Dragonfly | 3 | 2.71 | 206 |
| **(264,11)-Optimal** | 3 | 2.50 | 422 |
| (264,11)-Dragonfly | 3 | 2.69 | 278 |

### 4.4.2. Simulation Results and Analysis

We simulated larger-scale topologies on the platform SimGrid (version 3.21) [59]. SimGrid provides versatile, accurate and scalable simulation of distributed applications, especially with SMPI API that enables simulation of unmodified MPI applications [59]. We configure SimGrid to approximate the settings and ping-pong test results of Taishan cluster, with dual-core CPU per host, 8 Gflops processing speed per core, gigabit bandwidth and 30 μs latency per link. Static shortest-path routing is implemented with full routing table calculated using the same algorithm as for the benchmarking cluster. We run the simulations on the SeaWulf cluster at Stony Brook University.

We select the benchmarks that require global communication: MPI_Alltoall, effective bandwidth, 1D FFTE, Graph 500 and NPB IS and FT. Because of the limited 128 GB RAM of SeaWulf nodes and long simulation runtime for large-scale topologies, we reduced the problem sizes for some benchmarks, namely, 64 KB and 512 KB as the unit message sizes for MPI_Alltoall, 1 MB maximum message size for effective bandwidth and Classes S and A for NPB IS. For Graph 500, due to implementation issues with SimGrid, we used a previous version 2.1.4 that only contains BFS test and reduced the test scale to 12. For high-radix topologies we excluded FFTE and NPB as in Section 4.3.

The simulation performance ratios to ring are plotted in Fig. for low-radix topologies of 256 nodes, with log scale on MPL. The simulation performance ratios of optimal topologies over Dragonfly topologies are listed in **Erro! Fonte de referência não encontrada.**. The suboptimal topologies are labeled as (N,k)-S and gold (or cyan) points indicate the data for degree 6 (or 8) clusters.

The simulation results reveal that for large-scale low-radix topologies, (256,k)-Suboptimal with low MPL have mostly prominent performance increase over other topologies with the same degree. Despite fluctuations in Graph 500 BFS (Fig.d) and NPB IS (Fig.e) due to limited problem sizes and thus less intensive communication, all the simulation performances show a strongly inversely proportional relation with respect to MPL. The performance gain of (256,8)-Suboptimal over (256,3)-Wagner is above 1000% in MPI_Alltoall (Fig.a**Erro! Fonte de referência não encontrada.**), 1D FFTE (Fig.c) and NPB FT (Fig.f). Again, tori show low performance partially due to network congestion caused by static shortest-path routing.

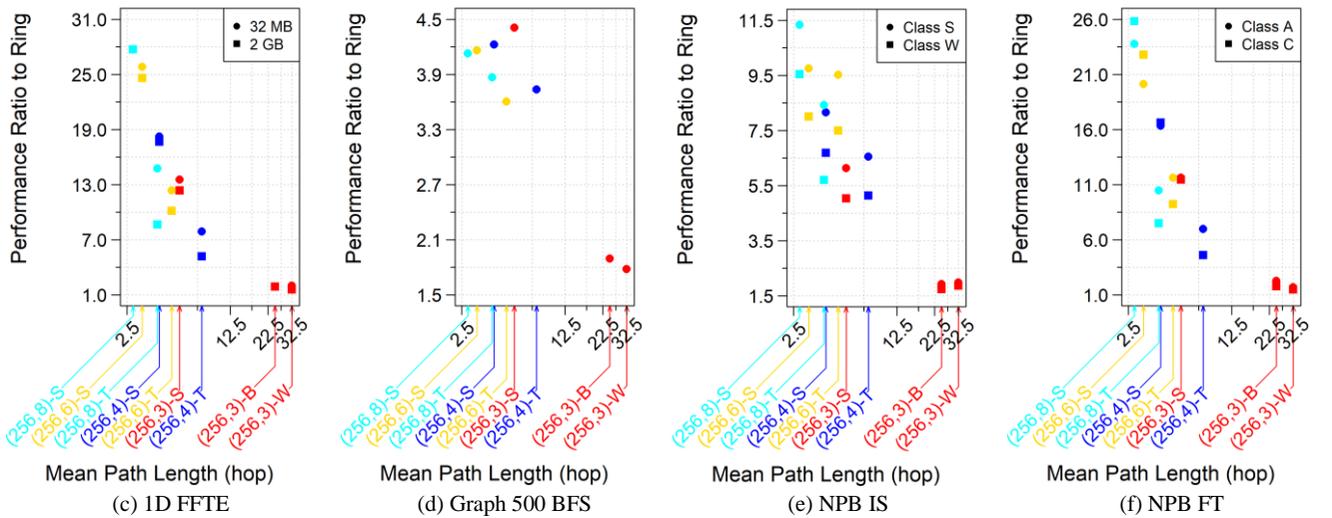

Fig. 10. Performance ratios on simulated MPI_Alltoall, effective bandwidth, 1D FFTE, Graph 500 BFS and NPB.

TABLE 6 Relative Performances of Larger-scale Optimal Topologies Over Corresponding Dragonfly Topologies

| Topology | b_eff | Graph 500 | MPI_Alltoall | | | |
|---|---|---|---|---|---|---|
| $(N,k)$ | | | BFS | 64 KB | 512 KB | |
| (252,11) | | | 1.14 | 0.97 | 1.92 | 2.57 |
| (264,11) | | | 0.92 | 1.05 | 1.72 | 2.29 |

For large-scale high-radix topologies, the optimal topologies show better performance especially in MPI_Alltoall, where (252,11)-Optimal has top performance increase of 92%/157% over Dragonfly for unit message sizes 64 KB/512 KB. For effective bandwidth and Graph 500, the optimal and dragonfly topologies have nearly the same performance with slight fluctuations, resulting from their equal diameter and relatively close MPL.

## 5 DISCUSSION AND CONCLUSION

We approached our heuristics that clusters of the same size with network topologies of minimal MPL will, in general, outperform those with higher MPL. That is done experimentally in small clusters with network topologies configured by optimal regular graphs. We built clusters of the same size with multiple topologies including Dragonfly, torus, Wagner, Bidiakis, Chvatal, and ring, and run a variety of benchmarks and applications. Our results show that the optimal network topologies, in general, deliver the highest performance. Our simulations for larger clusters confirm the same observations. The minimum MPL graphs were constructed using our algorithm based on simulated annealing that uses a reduced search space based on girth restrictions and symmetry requirements for the graph. This method is general, being applied well for the search of low and high-radix topologies as demonstrated by our search on optimal network topologies to compare with the Dragonfly. Based on our experimental and simulation results, it is fair to assume that the optimal network topologies will help supercomputer architects to maximize communication performance and peak processing speeds in a cluster.

Our results for applications with high communication to computation ratios, namely, MPI_Alltoall-based tests, effective bandwidth, 1D FFTE, Graph 500, and NPB IS and FT, indicate the strong influence of MPL on the clusters' performance. This proves the importance of network topologies with optimized MPL for speeding up processing and encourages designing clusters of 64, or 128, nodes for investigating the gains in peak processing speeds accordingly with network topology. Note that our approach might also be used on the design of the communicating circuitry of multicore processors or on the design of optimal low-radix topologies for DCNs.

Designing optimal network topologies for minimal MPL with additional requirement of symmetrical structure is demonstrated to be important for both enhancing performance and ensuring engineering feasibility. Such an approach is useful as it enables a better use of the available hardware while adding minimal costs: the time and energy for computational search of the optimal topology for a regular graph of a given size and node degree. That feature indicates the necessity of developing mathematical tools for minimizing the computer search time or, in an ideal scenario, finding optimal graph topologies by analytic calculations. Currently, the parallel exhaustive search for (32,3)-Optimal without girth constraint goes through ~$10^{13}$ graphs and took about one week on thousands of Sunway BlueLight cores. That amount of time is greatly reduced if we consider the symmetries and obtain sub-optimal graphs as done for the 256 nodes graphs.

The linear relation between the distance and latency matrices for, respectively, the graph and the networks demonstrate the usefulness our mathematically driven design. Tables TABLE 1, TABLE 2, TABLE 4 and **Erro! Fonte de referência não encontrada.** show the properties of the networks that we have evaluated in our work and (sub)optimal graphs with the additional requirement of symmetry, also have maximized (minimized) bisection width (diameter). Those two quantities also help on enhancing the cluster's performance as widely known by supercomputer and DCNs architects. Therefore, our approach provides an additional layer of optimization of a cluster's performance and points towards the necessity of construction of a theoretical body enabling to predict the clusters performance accordingly with its network topology.

### ACKNOWLEDGMENT

This work is supported by Key Research and Development Program of Shandong Province of China (2015GGX101028), National Natural Science Foundation of China (11674264), Science Foundation for Youths of Shandong Academy of Sciences of China (No. 2014QN010) and Special Fund for Overseas High-level Talents of Shandong Province of China. AFR thanks CAPES Process n. 88881.062174/2014-0 and Tinker Visiting Professorship at CLAS-University of Chicago. The authors thank the following individuals for their contributions to this project at different stages on various aspects: Y. Hu, X. Li, C. Chen, E. Raut, H. Fan, K. Wu, C. Han, Z. Ye, L. Zhang, D. McLoone, L. Pei, J. Han, H. Zheng, T. Mayer, and J. Mitchell et al.

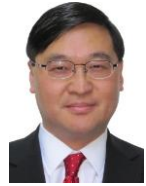

**Yuefan Deng** received the BS (1983) in physics from Nankai University and PhD (1989) from Columbia University. He is currently a professor of applied mathematics at Stony Brook University. He has built 2nd generation supercomputer for QCD at Columbia (in 1985), the Nankai Stars (2004) and RedNeurons (2007) Supercomputers. He holds 13 US patents on interconnection topologies and cloud resource sharing.

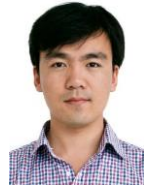

**Meng Guo** received the BS and PhD in physics from Shandong University in 2005 and 2012, respectively. He is currently a research scientist at National Supercomputer Center in Jinan, China. He has 10 years of experience on computational physics and high-performance computing.

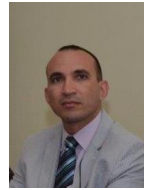

**Alexandre F. Ramos** received BS (2001) in physics from Federal University of São Carlos and and PhD (2008) in Physics from University of São Paulo. He is currently a professor of physics at University of São Paulo with 10 years of experience on application of theoretical methods of physics in biology and HPC research.

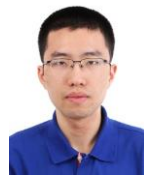

**Xiaolong Huang** received the BS degree in mathematics and applied mathematics from Zhejiang University in 2011. He is currently a PhD student in the Applied Mathematics and Statistics Department at Stony Brook University.

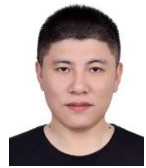

**Zhipeng Xu** received the MS degree in system theory from Guangxi Normal University in 2017. He is currently a PhD student in School of Data and Computer Science at Sun Yat-sen University and a visiting student at Stony Brook University.

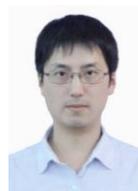

**Weifeng Liu** received the BS and PhD degrees in software engineering and computer architecture from Shandong University in 2010 and 2016 respectively. He is now a lecturer in the University of Jinan. He is working on MPI optimization and NoC designing for many years.